\documentclass[iop]{emulateapj}
\pdfoutput=1 
\usepackage{amsmath,amstext}
\usepackage[T1]{fontenc}
\usepackage{apjfonts}
\usepackage{hyperref}
\usepackage[figure,figure*]{hypcap}

\usepackage{color}

\shorttitle{Where are LIGO's big black holes?}
\shortauthors{Fishbach \& Holz}

\begin{document}

\title{Where are LIGO's big black holes?}

\author{Maya Fishbach}
\affiliation{Department of Astronomy and Astrophysics, University of Chicago, Chicago, IL 60637, USA}
\author{Daniel E. Holz}
\affiliation{Enrico Fermi Institute, Department of Physics, Department of Astronomy and Astrophysics,\\and Kavli Institute for Cosmological Physics, University of Chicago, Chicago, IL 60637, USA}

\begin{abstract}
In LIGO's O1 and O2 observational runs, the detectors were sensitive to stellar mass binary black hole coalescences with component masses up to $100\,M_\odot$, with binaries with primary masses above $40\,M_\odot$ representing $\gtrsim90\%$ of the total accessible sensitive volume. Nonetheless, of the 5.9 detections (GW150914, LVT151012, GW151226, GW170104, GW170608, GW170814) reported by LIGO-Virgo, the most massive binary detected was GW150914 with a primary component mass of $\sim36\,M_\odot$, far below the detection mass limit. Furthermore, there are theoretical arguments in favor of an upper mass gap, predicting an absence of black holes in the mass range $50\lesssim M\lesssim135\,M_\odot$.
We argue that the absence of detected binary systems with component masses heavier than $\sim40\,M_\odot$ may be preliminary evidence for this upper mass gap. By allowing for the presence of a mass gap, we find weaker constraints on the shape of the underlying mass distribution of binary black holes. We fit a power-law distribution with an upper mass cutoff to real and simulated BBH mass measurements, finding that the first 3.9 BBHs favor shallow power law slopes $\alpha \lesssim 3$ and an upper mass cutoff $M_\mathrm{max} \sim 40\,M_\odot$. This inferred distribution is entirely consistent with the two recently reported detections, GW170608 and GW170814.
We show that with $\sim10$ additional LIGO-Virgo BBH detections, fitting the BH mass distribution will provide strong evidence for an upper mass gap if one exists.
\end{abstract}

\section{Introduction}
One of the most fundamental quantities in gravitational-wave astrophysics is the mass distribution of stellar-mass black holes (BHs). Characterizing this distribution in merging binary systems is crucial to understanding stellar evolution, supernova physics, and the formation of compact binary systems. Prior to the first gravitational-wave (GW) detections of binary black holes (BBHs), the sample of $\sim 20$ BHs in X-ray binary systems was used to infer the BH mass distribution \citep{Ozel:2010,Farr:2011}, providing strong evidence for the existence of a mass gap between the heaviest neutron star (NS) ($\sim 2$--$3\,M_{\odot}$) and the lightest BH ($\sim4$--$5\,M_{\odot}$) \citep[but see also][]{Kreidberg:2012}. The presence of a mass gap between NSs and BHs has critical implications for supernova explosion theory \citep{Belczynski:2012}, and there are several proposed methods to probe this mass gap with gravitational-wave observations of compact binaries \citep{Littenberg:2015,Mandel:2017,Kovetz:2017}. In addition to the low-mass gap, supernova theory suggests that pulsational pair-instability supernovae (PPISN) \citep{HegerWoosley} and pair instability supernovae (PISN) \citep{1964ApJS....9..201F,1967ApJ...150..131R,Bond} lead to a second mass gap between $\sim 50$ and $135\,M_\odot$ for BHs formed from stellar core collapse \citep{Belczynski:2014,Marchant:2016,Belczynski:2016,Woosley:2017,Spera:2017}. Although studies of the lower mass gap have to wait for many more binary detections, because LIGO's sensitivity is almost 500 times greater for 50--$50\,M_\odot$ mergers than 3--$3\,M_\odot$ mergers, the existing data already begins to probe the upper mass gap.

For the first four BBH detections (GW150914, LVT151012, GW151226, and GW170104), the LIGO-Virgo collaboration fit the BBH mass distribution with a power law parametrization on the primary BBH mass, $m_1$, \citep{Abbott:O1BBH,Abbott:GW170104} inspired by the stellar initial mass function (IMF) \citep{Salpeter,Kroupa}. Specifically, \citet{Abbott:O1BBH} use the following one-parameter power law to model the distribution of primary component BH masses:
\begin{equation}
\label{Eq:LIGOpm1}
p\left( m_1 \mid \alpha \right) \propto m_1^{-\alpha},
\end{equation}
where $M_\mathrm{min} < m_1 < M_\mathrm{max}$. The mass ratio between component BHs, $q \equiv {m_2}/{m_1} \leq 1$, is assumed to be uniformly distributed in the allowed range $M_\mathrm{min}/m_1 \leq q \leq \mathrm{min}(M_\mathrm{tot,max}/m_1-1,1)$. Thus, the marginal distribution of the secondary component mass, $m_2$, is given by:
\begin{equation}
\label{Eq:LIGOpm2}
p\left( m_2 \mid m_1 \right) \propto \frac{1}{\mathrm{min}(m_1,M_\mathrm{tot,max}-m_1) - M_\mathrm{min}},
\end{equation}
and therefore the joint mass distribution is:
\begin{equation}
\label{Eq:LIGOpl}
p\left(m_1, m_2 \mid \alpha \right) \propto \frac{m_1^{-\alpha}}{\mathrm{min}(m_1,M_\mathrm{tot,max}-m_1) - M_\mathrm{min}}.
\end{equation}
The only free parameter in this assumed mass distribution is the power law slope, $\alpha$. The minimum BH mass, $M_\mathrm{min}$, is fixed at $M_\mathrm{min} = 5\,M_\odot$ and the maximum mass, $M_\mathrm{max}$, is fixed at $M_\mathrm{max} = 100\,M_\odot - M_\mathrm{min}$. Meanwhile, the total BBH mass, $M_\mathrm{tot} = m_1+m_2$,  is also restricted: $M_\mathrm{tot} \leq M_\mathrm{tot,max} = 100\,M_\odot$. (Enforcing $M_\mathrm{tot,max} = 100\,M_\odot$ causes a break in the power law at $50\,M_\odot$.)
The choice for $M_\mathrm{min}$ is motivated by the empirical lower mass gap, while $M_\mathrm{max}$ and $M_\mathrm{tot,max}$ are set by the stellar binary matched-filter search, which defines stellar-mass BBHs as those with source-frame total masses $m_1 + m_2  \leq 100\,M_\odot$ \citep{Cannon:2012,Usman:2016,Abbott:search}. However, LIGO is in principle sensitive to heavier BBHs \citep{Abbott:IMBH}. BBHs with detector-frame total masses up to $600\,M_\odot$ can be detected via matched-filtering by the intermediate mass black hole (IMBH) modeled search \citep{Nitz2017,Messick2017,DalCanton2017}, and IMBHs of even higher mass can be detected by the unmodeled transient search \citep{Klimenko2008,Abbott:burst}.

We observe that a key assumption of the distribution in Eq. \ref{Eq:LIGOpl} is that BHs in merging binaries follow the same mass distribution from $5\,M_\odot$ to at least $50\,M_\odot$, and that there exist BHs as heavy as $95\,M_\odot$. Meanwhile, LIGO is extremely sensitive to heavy BBHs. The first-order post-Newtonian approximation predicts that for low mass BBHs and a Euclidean universe, the spacetime volume, $VT$, to which LIGO can detect a BBH merger of a fixed mass ratio increases with its primary component mass, $m_1$, roughly as $VT \propto m_1^{5/2}$ . In the following section, we find that when accounting for cosmology and taking BBHs over the entire range $10\,M_\odot < M_\mathrm{tot} < 100\,M_\odot$, it is still a good approximation to take $VT \propto m_1^k$, with $k \sim 2.2$. This means that if the BBH mass distribution follows a power law with slope $\alpha$ as in Eq.~\ref{Eq:LIGOpl}, we expect the mass distribution among detected BHs to follow $m_1^{-\alpha+2.2}$. For a Salpeter IMF ($\alpha=2.35$), this implies an almost flat {\em detected}\/ distribution of binary black hole masses. Thus, the absence of heavy BBHs in the data quickly indicates either that the mass distribution declines steeply towards high masses ($\alpha \gg 2.2$), or that an upper mass gap sharply cuts off the mass distribution.

In this Letter we show that we can start to distinguish between these two scenarios with the first four LIGO BBH detections (including LVT151012, which has an $87\%$ probability of being astrophysical) \citep{Abbott:O1BBH,Abbott:GW170104}. Using simulated BBH detections, we demonstrate that if an existing mass gap is not accounted for, the non-detection of heavy stellar mass BHs will quickly bias the power law fit to distributions which are erroneously too steep. However, by including a maximum BH mass as a free parameter in the analysis, we can simultaneously infer the shape of the mass distribution and the location of a mass gap, if present. We carry out this analysis for the first four BBHs as well as for simulated BBH detections. We find that there is already evidence for an upper mass cutoff at $\sim40\,M_\odot$ from the first four detections, a conclusion that is further supported by the two recently reported BBH detections (GW170608 and GW170814) \citep{Abbott:GW170814, Abbott:GW170608}. We show that with $\mathcal{O}(10)$ additional detections, the presence and location of the bottom edge of the mass gap will be highly constrained.

\begin{figure}
\label{fig:VT_Mtot}
\includegraphics[width=0.5\textwidth]{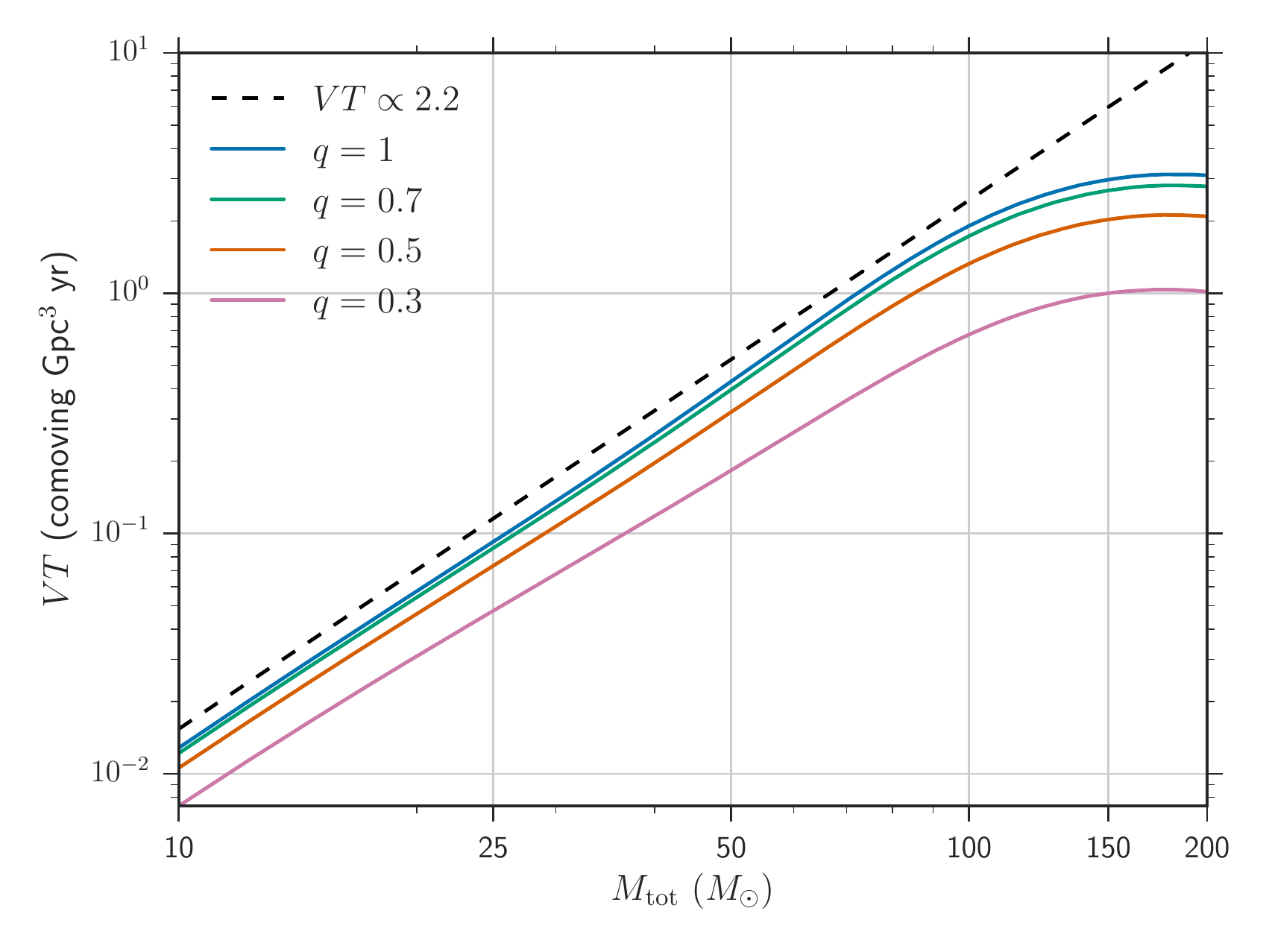}
\caption{Sensitive redshifted spacetime volume, $VT$, of the LIGO detectors in O1 and O2 as a function of BBH total mass, $M_\mathrm{tot}$, and mass ratio, $q$, calculated under the semi-analytic approximation described in the text for one year of observation. We find that $VT \propto m_1^{2.2}$ over the range $10\,M_\odot \leq M_\mathrm{tot} \leq 100 \,M_\odot$.}
\end{figure}

\section{Sensitive volume}
\label{section:volume}
As we noted in the previous section, for a given mass ratio, the sensitivity of the LIGO-Virgo search scales with primary component mass roughly as ${m_1}^{2.2}$. We characterize the sensitivity by the redshifted spacetime volume, $VT$, for which a given search is sensitive to a BBH system of given masses. If we assume that the rate of BBH coalescences is uniform in comoving volume and source-frame time and neglect the effects of BH spin on the detectability of a source, $VT$ depends only on the power spectral density (PSD) curve characterizing the detectors and the BBH component masses \citep{Abbott:rates}. Under these assumptions, $VT$ is given by:
\begin{equation}
\label{eq:VTm1m2}
VT(m_1,m_2) = T \int dz \frac{dV_c}{dz} \frac{1}{1+z} f(z,m_1,m_2),
\end{equation}
where $T$ is the search time, $V_c$ is the comoving volume, and $0 < f(z,m_1,m_2) < 1$ is the probability that a BBH system of masses $m_1$, $m_2$ at redshift $z$ will be detected. We adopt the cosmological parameters from \citet{Planck:2015} throughout the calculation. To calculate the detection probability, $f(z,m_1,m_2)$, we use the semi-analytic approximation from \citet{Abbott:rates}. Taking the PSD function corresponding to the early aLIGO high-sensitivity scenario in \citet{Abbott:OS} (a good approximation to the PSD during the first and second aLIGO observing runs), we calculate the optimal matched-filter signal-to-noise ratio (SNR) of a BBH with component masses $m_1$ and $m_2$ and zero spins located at redshift $z$. The optimal SNR, $\rho_\mathrm{opt}$, corresponds to a face-on source that is directly overhead to a single detector. We then generate random angular factors, $0 < w < 1$, from a single-detector antenna power pattern, assuming that binaries are distributed uniformly on the sky with isotropic inclination vectors \citep{FC:93,DominikIII,Belczynski:rates}. The angular factor, $w$, characterizes the response of a detector to a source at a given sky location and orientation (so that $w$ = 1 for an overhead, face-on source). From the distribution of angular factors, $w$, we assign a distribution of SNRs, $\rho = w \rho_\mathrm{opt}$, for each source with parameters $(m_1, m_2, z)$. Out of this distribution of SNRs, the fraction that exceed the single-detector threshold $\rho > 8$, roughly corresponding to a network threshold $\rho > 12$, is taken to be the detection probability $f(z, m_1, m_2)$. This semi-analytic calculation for $f(z,m_1,m_2)$ neglects the effects of non-Gaussian noise, which tends to raise the SNR detection threshold for binaries of very high mass. However, it remains a good approximation for stellar-mass binaries of total masses up to at least $M_\mathrm{tot}=100\,M_\odot$ and possibly higher \citep{Abbott:rates, Abbott:IMBH}. 

The expected sensitive redshifted spacetime volume as a function of total mass is shown in Fig.~\ref{fig:VT_Mtot}, calculated for one year of observation ($T$ = 1 year) at the O1-O2 LIGO sensitivity. For example, we note that in O1 and O2 the LIGO detectors probed a volume roughly seven times larger for 75--$75\,M_\odot$ binaries as compared to 25--$25\,M_\odot$ binaries. In particular, since $m_1 = M_\mathrm{tot}/(1+q)$ for a fixed mass ratio $q$, we can see from Fig.~\ref{fig:VT_Mtot} that $VT$ scales approximately as $m_1^k$ with $k \sim 2.2$ for $M_\mathrm{tot} \lesssim 100\,M_\odot$.

To calculate the sensitivity to a population of BBHs, the relevant quantity is the population-averaged spacetime volume, $\langle VT \rangle$. If we know the distribution of masses across the population of BBHs, $p_\mathrm{pop}(m_1,m_2)$, assuming negligible spins and a constant comoving merger rate, we can calculate the population-averaged sensitive spacetime volume \citep[Eq. 15 in][]{Abbott:rates}:
\begin{equation}
\label{eq:effVT}
\langle VT \rangle = \int \int VT(m_1,m_2)p_\mathrm{pop}(m_1,m_2)\,dm_1 dm_2,
\end{equation}
where the first integral is over $M_\mathrm{min} < m_1 < M_\mathrm{max}$ and the second integral is over $M_\mathrm{min} < m_2 < \mathrm{min}(m_1, \ M_\mathrm{max,tot}-m_1)$.
$\langle VT \rangle$ relates the specific merger rate, $R$, to the expected number, $\Lambda$, of BBH signals in a given detection period \citep{Abbott:rates}:
\begin{equation}
\Lambda = R\langle VT \rangle.
\end{equation}
The number of BBH detections, $n$, follows a Poisson process with mean $\Lambda$. To explore the existence of a high mass gap in Section~\ref{section:resultsVTratio}, we compare the expected number of low mass BBH signals, $\Lambda_\mathrm{low} = R \langle VT \rangle_\mathrm{low}$, to the expected number of high mass BBH signals, $\Lambda_\mathrm{high} = R \langle VT \rangle_\mathrm{high}$, for different power-law populations, where low (high) mass is defined by the primary component mass $m_1 \leq M_\mathrm{cutoff}$ ($m_1 > M_\mathrm{cutoff}$). We define:
\begin{equation}
\label{eq:VTratio}
\frac{1}{r} \equiv \frac{\Lambda_\mathrm{low}}{\Lambda_\mathrm{high}} = \frac{\langle VT \rangle_\mathrm{low}}{\langle VT \rangle_\mathrm{high}},
\end{equation}
where
\begin{equation}
\begin{split}
\label{eq:VTlowhigh}
\langle VT \rangle_\mathrm{low} &\equiv \int \int_{M_\mathrm{min}}^{M_\mathrm{cutoff}} VT(m_1,m_2)p_\mathrm{pop}(m_1,m_2)\, dm_1 dm_2 \\
\langle VT \rangle_\mathrm{high} &\equiv \int \int_{M_\mathrm{cutoff}}^{M_\mathrm{max}} VT(m_1,m_2)p_\mathrm{pop}(m_1,m_2)\, dm_1 dm_2.
\end{split}
\end{equation}
The integration limits on the $m_2$ integral in Eq.~\ref{eq:VTlowhigh} are identical to those in Eq.~\ref{eq:effVT} so that the total $ \langle VT \rangle = \langle VT \rangle_\mathrm{low} + \langle VT \rangle_\mathrm{high}$.
We can then compute the probability of detecting $n_\mathrm{high}$ BBHs with primary component mass $m_1 > M_\mathrm{cutoff}$, given that we have detected $n_\mathrm{low}$ BBHs with primary component mass $m_1 < M_\mathrm{cutoff}$. (We ignore mass measurement uncertainties which may prevent us from definitively assigning a BBH to either the low or high mass class.) This probability is given by:
\begin{equation}
\label{eq:pn2}
\begin{split}
&p\left( n_\mathrm{high} \mid n_\mathrm{low} \right) =
\int_0 ^\infty \int_0 ^\infty p\left( n_\mathrm{high}, \Lambda_\mathrm{high}, \Lambda_\mathrm{low} \mid n_\mathrm{low} \right) d \Lambda_\mathrm{high} d \Lambda_\mathrm{low}
\\
&= \int \int p\left( n_\mathrm{high} \mid \Lambda_\mathrm{high} \right) p \left( \Lambda_\mathrm{high} \mid \Lambda_\mathrm{low} \right) p \left( \Lambda_\mathrm{low} \mid n_\mathrm{low} \right) d \Lambda_\mathrm{high} d \Lambda_\mathrm{low} \\
& = \int \int p\left( n_\mathrm{high} \mid \Lambda_\mathrm{high} \right) \delta \left( \Lambda_\mathrm{high} - r \Lambda_\mathrm{low} \right) p \left( \Lambda_\mathrm{low} \mid n_\mathrm{low} \right) d \Lambda_\mathrm{high} d \Lambda_\mathrm{low} \\
& = \int p\left( n_\mathrm{high} \mid r\Lambda_\mathrm{low} \right) p \left( \Lambda_\mathrm{low} \mid n_\mathrm{low} \right) d \Lambda_\mathrm{low} \\
&\propto \int p\left( n_\mathrm{high} \mid r\Lambda_\mathrm{low} \right) p \left(  n_\mathrm{low} \mid \Lambda_\mathrm{low} \right) p_0 \left( \Lambda_\mathrm{low} \right) d \Lambda_\mathrm{low},
\end{split}
\end{equation}
where in the third line we used the definition of $r$ given by Eq.~\ref{eq:VTlowhigh} and in the last line we used Bayes's theorem. In Eq.~\ref{eq:pn2} terms like $p\left( n \mid \Lambda \right)$ denote the Poisson probability of $n$ with mean $\Lambda$. We take the prior $p_0 \left( \Lambda_\mathrm{low} \right)$ to be the Jeffrey's prior:
\begin{equation}
p_0 \left( \Lambda_\mathrm{low} \right) \propto \frac{1}{\sqrt{\Lambda_\mathrm{low}}}.
\end{equation}
We will return to Eq.~\ref{eq:pn2} in Section~\ref{section:resultsVTratio}.

\section{Fitting the Mass Distribution}
Our goal is to jointly infer the shape of the BBH mass distribution along with the lower edge of a potential mass gap, $M_\mathrm{max}$.
We therefore follow \citet{Abbott:O1BBH} in fitting a power-law mass distribution to gravitational-wave BBH mass measurements, but we add the maximum BH mass, $M_\mathrm{max}$, as a free parameter. We leave the minimum BH mass, $M_\mathrm{min}$, fixed at $M_\mathrm{min} = 5\,M_\odot$. Thus, we consider a two-parameter mass distribution:
\begin{equation}
\label{eq:pl_cutoff}
p\left( m_1, m_2 \mid \alpha, M_\mathrm{max} \right) \propto  \frac{m_1^{-\alpha} \ \mathcal{H}(M_\mathrm{max} - m_1)}{\min(m_1,M_\mathrm{tot,max}-m_1) - M_\mathrm{min}},
\end{equation}
where $\mathcal{H}$ is the Heaviside step function that enforces a cutoff in the distribution at $m_1 = M_\mathrm{max}$. For consistency with the LIGO definition of a stellar mass BH, we restrict $M_\mathrm{max} \leq 100\,M_\odot$ throughout. Furthermore, as in the LIGO collaboration's analysis, we enforce $m_1 + m_2 \leq M_\mathrm{tot,max}$, which provides an additional constraint for $M_\mathrm{max} > \frac{1}{2} M_\mathrm{tot,max}$. The LIGO collaboration fixes $M_\mathrm{tot,max} = 100\,M_\odot$ and $M_\mathrm{max} = 100\,M_\odot - M_\mathrm{min}$, as this is the definition of a stellar-mass BBH set by the search. This choice corresponds to one of the following assumptions: either (a) BBHs with total source-frame masses $M_\mathrm{tot} > 100\,M_\odot$ do not exist as part of the population of stellar-mass BBHs or (b) LIGO is not sensitive to BBHs with total source-frame masses $M_\mathrm{tot} > 100\,M_\odot$, so we cannot constrain their existence. (In the absence of detections with $M_\mathrm{tot,max} > 100\,M_\odot$, setting $M_\mathrm{tot,max} \leq 100\,M_\odot$ in the population model, Eq.~\ref{eq:pl_cutoff}, is equivalent to assuming that the sensitivity vanishes for binaries with $M_\mathrm{tot} > 100\,M_\odot$.) Assumption (a) may not be well-motivated, as population-synthesis models that predict stellar BBHs with component masses $M_\mathrm{max} \sim 50-100\,M_\odot$ tend to allow $M_\mathrm{tot, max} \sim 2 M_\mathrm{max}$ \citep{Belczynski:2016a, Eldridge:2016}. Assumption (b) can also be questioned, as LIGO is sensitive to BBHs with detector-frame total masses up to at least $600\,M_\odot$ in the IMBH matched-filter search \citep{Abbott:IMBH}. However, the sensitivity may be lower than expected for very high mass BBHs due to non-stationary instrumental noise \citep{Slutsky2010} and the absence of precessing and higher-order mode template waveforms, which leads to worse matches between signal and template for very high mass BBHs in the matched-filter search \citep{Capano2014,Bustillo2017,Bustillo2017a}. If we had an accurate model of $VT(m_1,m_2)$ across the mass range $5 < m_1, m_2 < 100\,M_\odot$ (by performing a large-scale injection campaign), we could set $M_\mathrm{tot,max} = 2M_\mathrm{max} \leq 200\,M_\odot$ in Eq.~\ref{eq:pl_cutoff}. However, because our calculation of $VT$ may be overestimating the sensitivity to binaries with $M_\mathrm{tot,max} > 100\,M_\odot$, when fitting Eq.~\ref{eq:pl_cutoff} in the following sections, we repeat the analysis once under the assumption that $M_\mathrm{tot,max} = \mathrm{min}(2M_\mathrm{max},100\,M_\odot)$ and once assuming that $M_\mathrm{tot,max} = 2M_\mathrm{max} \leq 200\,M_\odot$. 

To extract the parameters of our assumed mass distribution (Eq.~\ref{eq:pl_cutoff}) from data, we use the same hierarchical Bayesian methods as Appendix D of \citet{Abbott:O1BBH}, further explained in \citet{Mandel:2016}.
While GW data is noisy and subject to selection effects, both the measurement uncertainties and selection effects are well-quantified. The selection effects refer to the mass-dependent detection efficiency. Under the assumptions of negligible BH spins and a uniform comoving merger rate, the detection efficiency is proportional to the sensitive spacetime volume $VT(m_1,m_2)$ as described in Section~\ref{section:volume} and \citet{Abbott:rates}. 

BBH masses are measured using the LALInference parameter-estimation pipeline, which calculates the posterior probability density function (PDF) of all parameters that govern the waveform given the data, $d_i$, from a BBH detection \citep{Veitch:2015}. For an individual system, measurements of $m_1$ and $m_2$ take the form of $\mathcal{O}(10,000)$ posterior samples drawn from the posterior PDF, $p \left( m_1, m_2 \mid d_i \right)$.
In the following section, we perform our analysis on published mass measurements from the first four BBHs as well as on simulated BBH measurements. We use the fact that the one-dimensional PDFs for the source-frame chirp mass and symmetric mass ratio are well-approximated by independent (uncorrelated) Gaussian distributions. 

For the first four BBH sources, GW150914, LVT151012, GW151226, GW170104, we approximate the source-frame chirp mass posterior PDF as a Gaussian with a mean and standard deviation given by the median and $90 \%$ credible intervals listed in Table~4 of \citet{Abbott:O1BBH} or Table~1 of \citet{Abbott:GW170104}. In the case that the $90 \%$ credible interval is slightly asymmetric about the median, we use the average to estimate the standard deviation. We likewise approximate the posterior PDF of the symmetric mass ratio, $\eta = q/(1+q)^2$, as a Gaussian truncated to the allowed range $[0,0.25]$, with a mean and standard deviation given by the entry for $q$ in the same tables. Using these approximate chirp mass and symmetric mass ratio distributions, we generate 25,000 posterior samples from the component mass posterior PDFs of each event.

For our set of simulated BBH detections, we generate a set of component masses from an underlying mass distribution. To each BBH system, we assign a redshift from a redshift distribution that is uniform in the merger-frame comoving volume. Given the simulated masses and redshift of each BBH, we randomly generate its single-detector SNR from the antenna power pattern, using the PSD corresponding to the early aLIGO high-sensitivity scenario (as described in Section~\ref{section:volume}). Out of this population, the set of detections are those simulated BBHs with a single-detector SNR satisfying $\rho > 8$. Given the true component masses and the SNR of each mock BBH detection, we produce realistic mass measurements by generating 5,000--10,000 posterior samples for the component masses following the prescription in Eq.~1 of \citet{Mandel:2017}. Given true values for the chirp mass, $\mathcal{M}^T$, symmetric mass ratio, $\eta^T$, and SNR, $\rho^T$, we draw chirp mass posterior samples from a Gaussian distribution centered at $\bar{\mathcal{M}}$ with standard deviation $\sigma_{\mathcal{M}}$ and symmetric mass ratio posterior samples from a Gaussian distribution centered at $\bar{\eta}$ with standard deviation $\sigma_\eta$. We only keep posterior samples with $0.01 \lesssim \eta \lesssim 0.25$. The variables $\bar{\mathcal{M}}$ and $\bar{\eta}$ are drawn from Gaussian distributions:
\begin{equation}
\begin{split}
\bar{\mathcal{M}} &\sim N(\mathcal{M}^T,\sigma_{\mathcal{M}}), \\
\bar{\eta} &\sim N(\eta^T,\sigma_{\eta}),
\end{split}
\end{equation}
where $\sigma_\mathcal{M}$, $\sigma_\eta$ scale inversely with the SNR, and are given in \cite{Mandel:2017}. 

Once we have samples from the posterior PDF, $p \left( m_1, m_2 \mid d_i \right)$, for each event (both real and simulated) and we have calculated the detection efficiency, $P_\mathrm{det}(m_1,m_2) \propto VT(m_1,m_2)$, we follow Appendix D in \citet{Abbott:O1BBH} to fit Eq.~\ref{eq:pl_cutoff}. The likelihood for a single BBH detection given the parameters of the mass distribution, $\alpha$ and $M_\mathrm{max}$, is given by:
\begin{equation}
\label{eq:likelihood_plmc}
\begin{split}
p \left( d_i \mid \alpha, M_\mathrm{max} \right) &\propto \frac{ \int p \left( d_i \mid m_1, m_2 \right) p \left( m_1, m_2 \mid \alpha, M_\mathrm{max} \right) dm_1 dm_2}{\beta \left( \alpha, M_\mathrm{max} \right)} \\
&\propto \frac{\langle p \left( m_1, m_2 \mid \alpha, M_\mathrm{max} \right) \rangle }{\beta \left( \alpha, M_\mathrm{max} \right)},
\end{split}
\end{equation}
where $\langle \ldots \rangle$ denotes an average over the $(m_1, m_2)$ posterior samples. This is valid because for each event, $p \left( d_i \mid m_1, m_2 \right) \propto p \left( m_1, m_2 \mid d_i \right)$, as the prior on $m_1, \ m_2$ is taken to be flat. Therefore, we can calculate the integral in the first line of Eq.~\ref{eq:likelihood_plmc} by taking the average of $p(m_1,m_2 \mid \alpha, M_\mathrm{max})$ over the mass posterior samples. Meanwhile, $\beta(\alpha, M_\mathrm{max})$ is defined as:
\begin{equation}
\beta(\alpha, M_\mathrm{max}) \equiv \int p\left( m_1, m_2 \mid \alpha, M_\mathrm{max} \right) VT(m_1,m_2)\,dm_1 dm_2.
\end{equation}
 The likelihood for the data across all events $\mathbf{d} = \lbrace d_i \rbrace$ is the product of the individual event likelihoods given by Eq.~\ref{eq:likelihood_plmc}.

Furthermore, if we fix $M_\mathrm{max}$ and assume a prior $p_0 \left( \alpha \mid M_\mathrm{max} \right)$, we can calculate the Bayesian evidence in favor of a given $M_\mathrm{max}$:
\begin{equation}
\label{eq:evidence}
p\left(d_i \mid M_\mathrm{max} \right) = \int p \left( d_i \mid \alpha, M_\mathrm{max} \right) p_0 \left( \alpha \mid M_\mathrm{max} \right) d \alpha.
\end{equation}
We can then calculate the Bayes factor between two power-law models that differ in their choice of $M_\mathrm{max}$. Recall that the default LIGO analysis fixes $M_\mathrm{max} = 100\,M_\odot - M_\mathrm{min}$. For a sample of $N$ detected BBHs (assumed to be independent), the cumulative Bayes factor $K(M_\mathrm{max},100\,M_\odot)$ between a power-law model that fixes $M_\mathrm{max} = M$ and one that fixes $M_\mathrm{max} = 100\,M_\odot$ is a product of the single-event evidence ratios:
\begin{equation}
\label{eq:BayesFactor}
K(M,100\,M_\odot) =
\prod_{i=1}^N
\frac{p\left(d_i \mid M_\mathrm{max} = M \right)}{p\left(d_i \mid M_\mathrm{max} = 100\,M_\odot \right)}.
\end{equation}
We calculate the cumulative Bayes factor $K(M,100\,M_\odot)$ in Section~\ref{section:resultsBF}. 

\section{Results}
\label{section:results}
\subsection{Non-detection of heavy BBHs}
\label{section:resultsVTratio}

\begin{figure}
\label{fig:VTratio-pl}
\includegraphics[width=0.5\textwidth]{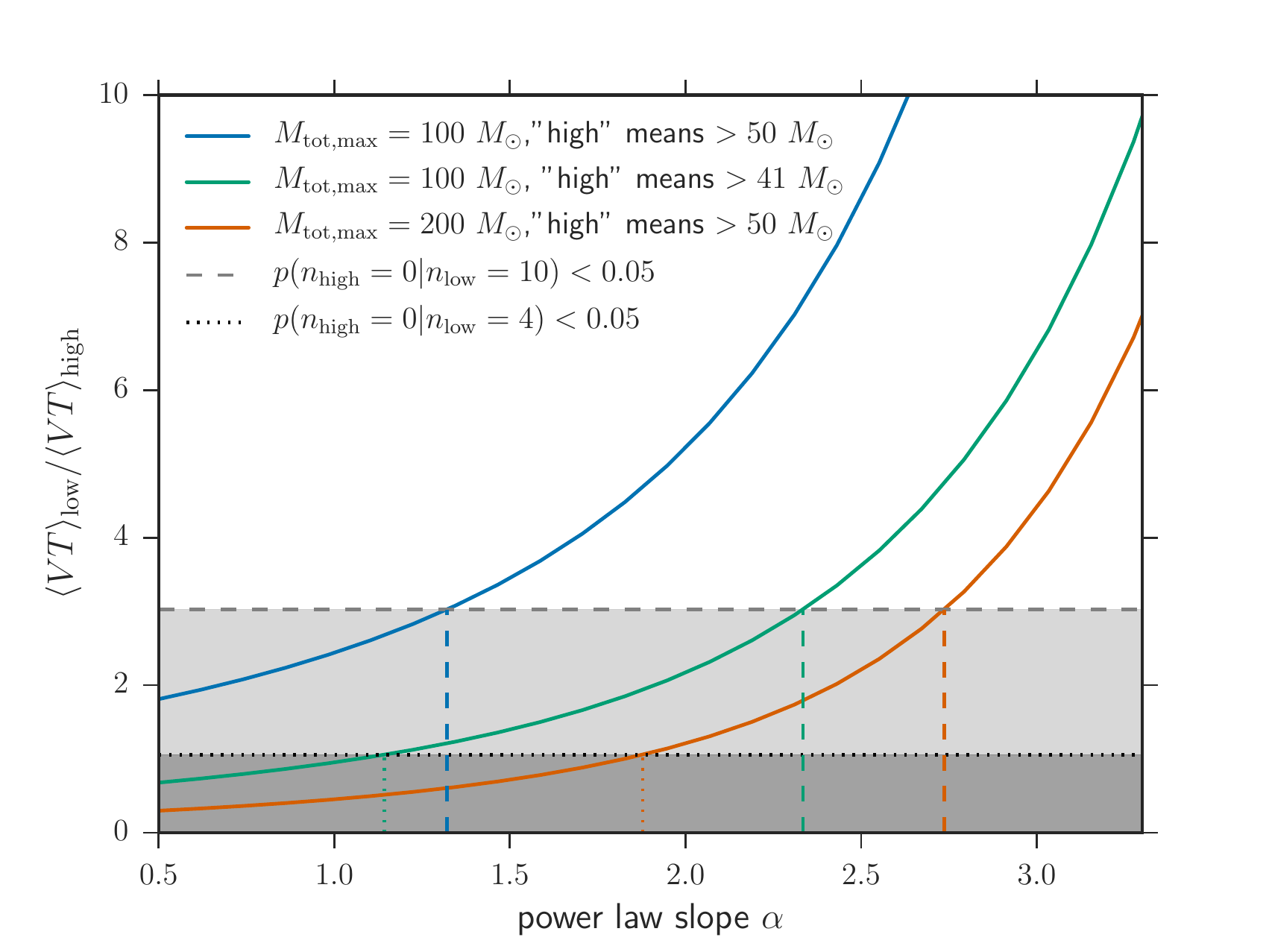}
\caption{Ratio ${\langle VT \rangle_\mathrm{low}}/{\langle VT \rangle_\mathrm{high}}$  of the expected number of``low" mass to ``high" mass BBH detections from an underlying population with power law slope $\alpha$ (solid curves). The blue and orange curves define low mass BBHs to have $m_1 < 50\,M_\odot$, while the green curve defines low mass as $m_1 < 40\,M_\odot$. The blue and green curves conservatively assume that the mass distribution and detector sensitivity extend only up to a total BBH mass of $100\,M_\odot$, while the orange curve assumes that the BBH population and sensitivity extend up to total masses of $200\,M_\odot$. The dashed (dotted) horizontal line corresponds to values of the ratio ${\langle VT \rangle_\mathrm{low}}/{\langle VT \rangle_\mathrm{high}}$ for which the probability of detecting no high mass BBHs and ten (four) low mass BBHs is less than $5 \%$. $VT$ ratios below this line correspond to values of $\alpha$ that lie to the left of the vertical colored dashed lines. With enough low mass BBH detections, shallow power law slopes (small positive values of $\alpha$) become inconsistent with the existence of high mass BBHs.}
\end{figure}

We first give an example of how the detection of only four BBHs with primary masses $m_1 \leq 40$--$50\,M_\odot$ is inconsistent with certain (possibly correct) power-law mass distributions unless a mass gap is imposed. For a given mass distribution, we can use Eq.~\ref{eq:pn2} to calculate the probability of not detecting any BBHs above a certain mass, $n_{\mathrm{high}}=0$, given that we have detected $n_\mathrm{low}$ BBHs below the cutoff mass. To do this, we must first compute the ratio $\langle VT \rangle _\mathrm{low} / \langle VT \rangle _\mathrm{high}$ as defined in Eq.~\ref{eq:VTratio}, and then apply Eq.~\ref{eq:pn2}. The results of this calculation for $p(n_\mathrm{high} = 0 \mid n_\mathrm{low} = 4)$ and $p(n_\mathrm{high} = 0 \mid n_\mathrm{low} = 10)$ are displayed in Fig.~\ref{fig:VTratio-pl}. We show the results for two choices of cutoff mass: $M_\mathrm{cutoff} = 41\,M_\odot$ (green curve) is motivated by the $95 \%$ credible upper limit on the primary mass of GW150914, the heaviest BBH detected, and $M_\mathrm{cutoff} = 50\,M_\odot$ (blue and orange curves) is motivated by PPISN and PISN supernova models, which predict a mass gap starting at $40$--$50\,M_\odot$ (depending also on details of binary evolution) \citep{Belczynski:2016,Woosley:2017}. We also vary the maximum total mass, $M_\mathrm{tot,max}$, of the "high mass" population between $100\,M_\odot$ (blue and green curves), which is currently the maximum total mass that the aLIGO search includes in the definition of a stellar mass BBH, to $200\,M_\odot$ (orange curve). We note that a power law with slope $\alpha = 1$ is the ``flat in log'' population that the LIGO-Virgo collaboration uses to compute the lower limits on the BBH merger rate, and we calculate that unless a mass gap is imposed, detecting four BBHs with primary masses $m_1 < 41\,M_\odot$ is inconsistent with this population at the $96\%$ level if we restrict $M_\mathrm{tot,max} = 100\,M_\odot$, or at the $> 99.9\%$ level if we assume that the BBH population and the detectors' sensitivity extends up to $M_\mathrm{tot,max} = 200\,M_\odot$. Furthermore, unless a mass gap is imposed, there is already some tension (inconsistency at the 93\% level) with the $\alpha = 2.35$ population that LIGO-Virgo uses to compute the upper rate limits if we assume $M_\mathrm{tot,max} = 200\,M_\odot$. If the BBH mass distribution has an upper cutoff at $40$--$50\,M_\odot$, the inferred merger rates calculated without assuming the cutoff would be $1.4$--$2.1$ times higher for the ``flat in log'' population and $1.1$--$1.4$ times higher for the $\alpha = 2.35$ population.

\subsection{Bayesian evidence in favor of mass gap}
\label{section:resultsBF}
\begin{figure}
\label{fig:alpha-c41}
\includegraphics[width=0.5\textwidth]{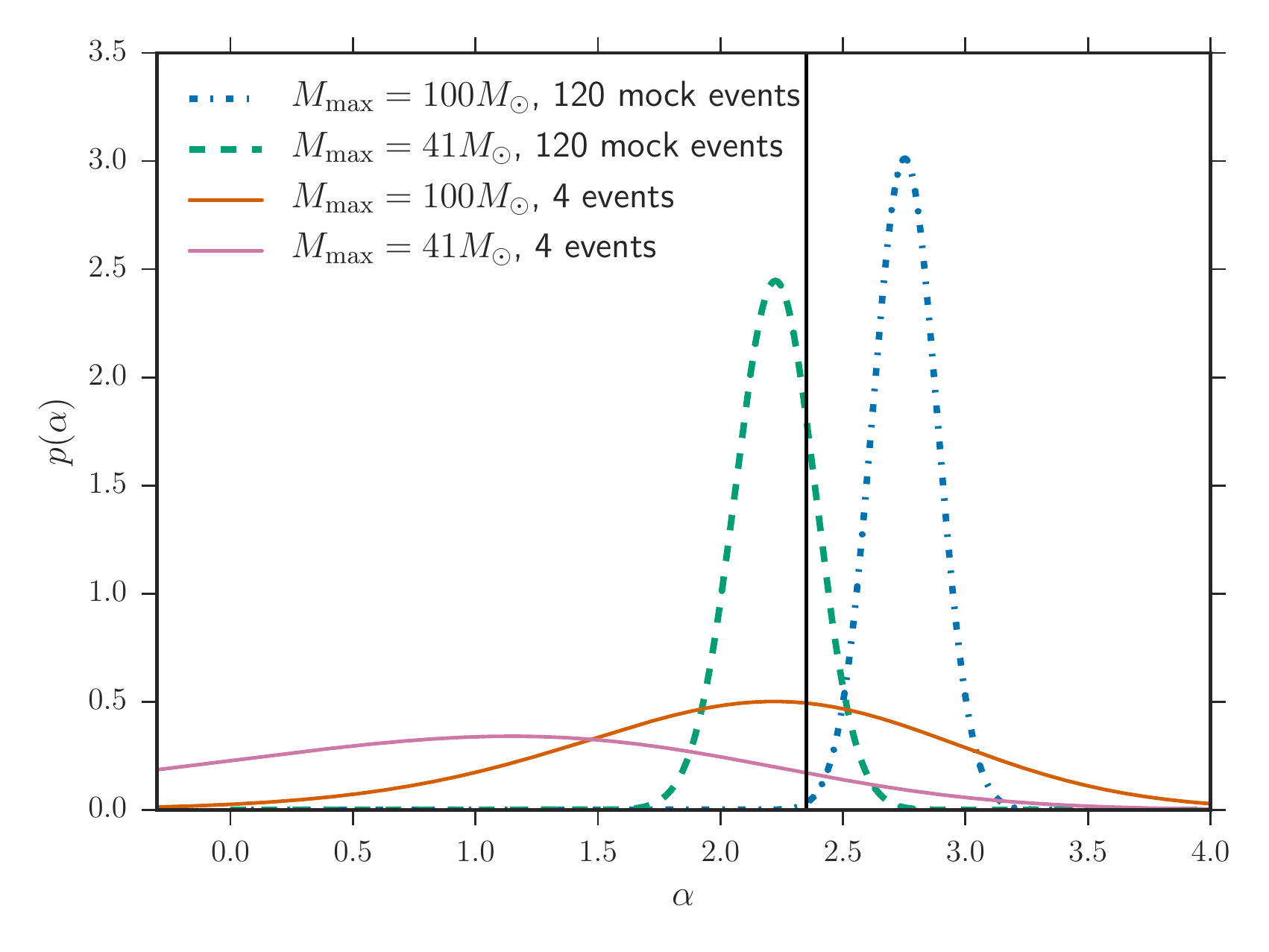}
\caption{Inferred likelihood for the power law slope of the mass distribution, $\alpha$, calculated for 120 mock observations from a $M_\mathrm{max} = 41\,M_\odot$, $\alpha = 2.35$ population (dashed and dotted curves) and the first four BBH detections (solid curves). The blue and orange curves correspond to the canonical LVC analysis in which the maximum mass of the BBH mass distribution is set to $M_\mathrm{max} = 95\,M_\odot$, while the pink and green curves correspond to a fixed maximum mass at $M_\mathrm{max} = 41\,M_\odot$. Neglecting to account for a high mass cutoff biases the power law inference to steep slopes. The solid black line at $\alpha = 2.35$ is the true slope of the simulated population, but gets ruled out by the canonical analysis.}
\end{figure}

We have seen that assuming a single power-law mass distribution over the entire mass range 5--$100\,M_\odot$ can rule out shallow power law slopes in the absence of detections with component masses $m_1 > 40$--$50\,M_\odot$. The absence of high mass detections will continue to push the inferred power law slope to steeper values unless we allow for an upper mass gap in the analysis. To study this point further, we simulate mock BBH mass measurements from a power-law population with slope $\alpha = 2.35$ and an upper mass cutoff at $M_\mathrm{max} = 41\,M_\odot$) (Eq.~\ref{eq:pl_cutoff}, but we follow the canonical analysis used by the LVC (see Eq.~\ref{Eq:LIGOpl}) and fix $M_\mathrm{max} = 95\,M_\odot$ and $M_\mathrm{tot,max} = 100\,M_\odot$ when inferring the power law slope. While the bias on the inferred slope $\alpha$ may be small with $\mathcal{O}(10)$ detections, with $\mathcal{O}(100)$ detections, the canonical analysis will rule out the correct power law slope (see Fig.~\ref{fig:alpha-c41}). (Although for hundreds of detections, a non-parametric fit to the mass distribution should be considered.) If we follow the canonical LIGO-Virgo analysis but set $M_\mathrm{max} = 100\,M_\odot$ and $M_\mathrm{tot,max} = 200\,M_\odot$ rather than $M_\mathrm{tot,max} = 100\,M_\odot$, the presence of a mass gap will bias the power law inference even more significantly, as the true population has $M_\mathrm{tot,max} = 82\,M_\odot$.
These results show that failing to account for an upper mass gap may lead to incorrect conclusions about the low mass distribution. While we demonstrated this for an assumed power law model, this caveat applies to any parametrized fit to the BBH mass distribution.

With the first four LIGO BBH detections, varying $M_\mathrm{max}$ does not drastically bias the inference on $\alpha$ when fitting Eq.~\ref{Eq:LIGOpl} (see the solid lines in Fig.~\ref{fig:alpha-c41}).
However, we can distinguish the model favored by the data by calculating the cumulative Bayes factor. Following Eqs.~\ref{eq:evidence}--\ref{eq:BayesFactor}, we calculate this factor between two single-parameter power law models with different fixed values of $M_\mathrm{max}$. We choose to compare two cutoff values, $41\,M_\odot$ (the $95\%$ upper limit on the heaviest component BH detected) and $100\,M_\odot$, and take the prior $p_0 \left(\alpha \mid M_\mathrm{max}\right)$ in Eq. \ref{eq:evidence} to be a top hat over the wide range $-2 < \alpha < 7$.

For the first four BBH detections,  $K(41\,M_\odot,100\,M_\odot) = 13$ if we assume the detectable BBH population only extends to $M_\mathrm{tot,max} = 100\,M_\odot$ (so that $M_\mathrm{max} = 100\,M_\odot$ is really $M_\mathrm{max} = 95\,M_\odot$). If we instead assume that the $M_\mathrm{max} = 100\,M_\odot$ population is fully detectable up to total binary masses of $M_\mathrm{tot,max} = 200\,M_\odot$, the Bayes factor increases to $K(41\,M_\odot,100\,M_\odot) = 90$, suggesting that there is already strong support for an upper mass cutoff at $M_\mathrm{max} \sim 40\,M_\odot$ over a cutoff at $M_\mathrm{max} \sim 100\,M_\odot$ within the assumed power law model \citep{KassRaftery}. The Bayes factor also depends on the choice of prior on $\alpha$. We choose to be relatively uninformative in our prior, excluding only very steeply declining mass distributions ($\alpha > 7$) and allowing for moderately upward sloping mass distributions ($-2 < \alpha < 0$), but it is clear from Fig.~\ref{fig:VTratio-pl} that a prior that favors large positive values of $\alpha$ (steeply declining power law slopes) will lower the evidence in favor of a mass cutoff $M_\mathrm{max} < 100\,M_\odot$, while placing greater prior support on low values of $\alpha$ (shallow or downward sloping power laws) will raise the evidence in favor of a mass cutoff. If we enforce $\alpha > 0$ in the prior in order to agree with other astrophysical mass distributions, the Bayes factors change to $K(41\,M_\odot,100\,M_\odot) = 5$ if we restrict $M_\mathrm{tot,max} \leq 100\,M_\odot$ or $K(41\,M_\odot,100\,M_\odot) = 21$ if we allow $M_\mathrm{tot,max} \leq 200\,M_\odot$. 

We anticipate that a set of ten BBH detections with primary component masses $m_1 \leq 41\,M_\odot$ will yield a Bayes factor $K(41\,M_\odot,100\,M_\odot) > 150$, providing very strong evidence for an upper mass gap. We assume that the underlying BBH population (and aLIGO's sensitivity) extends to total masses of $2 M_\mathrm{max}$ in either case, so $M_\mathrm{tot,max} \leq 200\,M_\odot$. We take a flat prior on $\alpha$ in the range $[-2, 7]$. With 191 events from a simulated $\alpha = 2.35$, $M_\mathrm{max} = 41\,M_\odot$ population, the single-event evidence ratios range from $K(41\,M_\odot, 100\,M_\odot) = 1.0$ to $K(41\,M_\odot,100\,M_\odot) = 16.9$, with a median of $K(41\,M_\odot, 100\,M_\odot) = 2.6$. With a subset of ten BBH detections from this population, we get $K(41\,M_\odot,100\,M_\odot) > 150$ in more than $99\%$ of cases. If we detect ten BBHs with primary component masses $m_1 \leq 50\,M_\odot$, we likewise expect very strong evidence for a mass cutoff, with $K(50\,M_\odot,100\,M_\odot) > 150$ more than $95\%$ of the time. The Bayes factor only compares two values of the mass cutoff; we fit for the value of $M_\mathrm{max}$ favored by a given set of detections in the following subsection~\ref{section:resultsjointfit}.

\subsection{Joint power law--maximum mass fit}
\label{section:resultsjointfit}
In this section we fit the two-parameter mass distribution of Eq.~\ref{eq:pl_cutoff}. We calculate the likelihood $p(\mathbf{d} \mid \alpha, M_\mathrm{max})$ as the product of individual event likelihoods in Eq.~\ref{eq:likelihood_plmc}. We take flat priors on $\alpha$ and $M_\mathrm{max}$, with $-2 \leq \alpha \leq 7$ as before and $M_\mathrm{max} \leq 100\,M_\odot$. The minimum allowed value of $M_\mathrm{max}$ for a given set of detections is set by the lower mass bound of the heaviest detected component BH. For simplicity, we take the lower mass bound to be the minimum posterior sample. The upper bound $M_\mathrm{max} \leq 100\,M_\odot$ is motivated by the LIGO stellar mass BBH search as well as by population synthesis studies, which usually predict that the BH mass distribution would extend to $80$--$130\,M_\odot$ were it not for a pair-instability mass gap \citep{Eldridge:2016, Belczynski:2016, Spera:2016}. We calculate the likelihood function on a $500 \times 100$ grid of $(\alpha, M_\mathrm{max})$ values in the allowed prior range, and verify that increasing the resolution of the $(\alpha, M_\mathrm{max})$ grid does not change our results. In fact, the resolution in the $M_\mathrm{max}$ dimension is limited by the finite number of posterior samples that are used to represent the component mass posterior PDFs for each event. To reduce these artificial discontinuities in the $M_\mathrm{max}$ dimension of the likelihood evaluation, we apply a two-dimensional smoothing spline before displaying the results.

\begin{figure*}
\includegraphics[width=\textwidth]{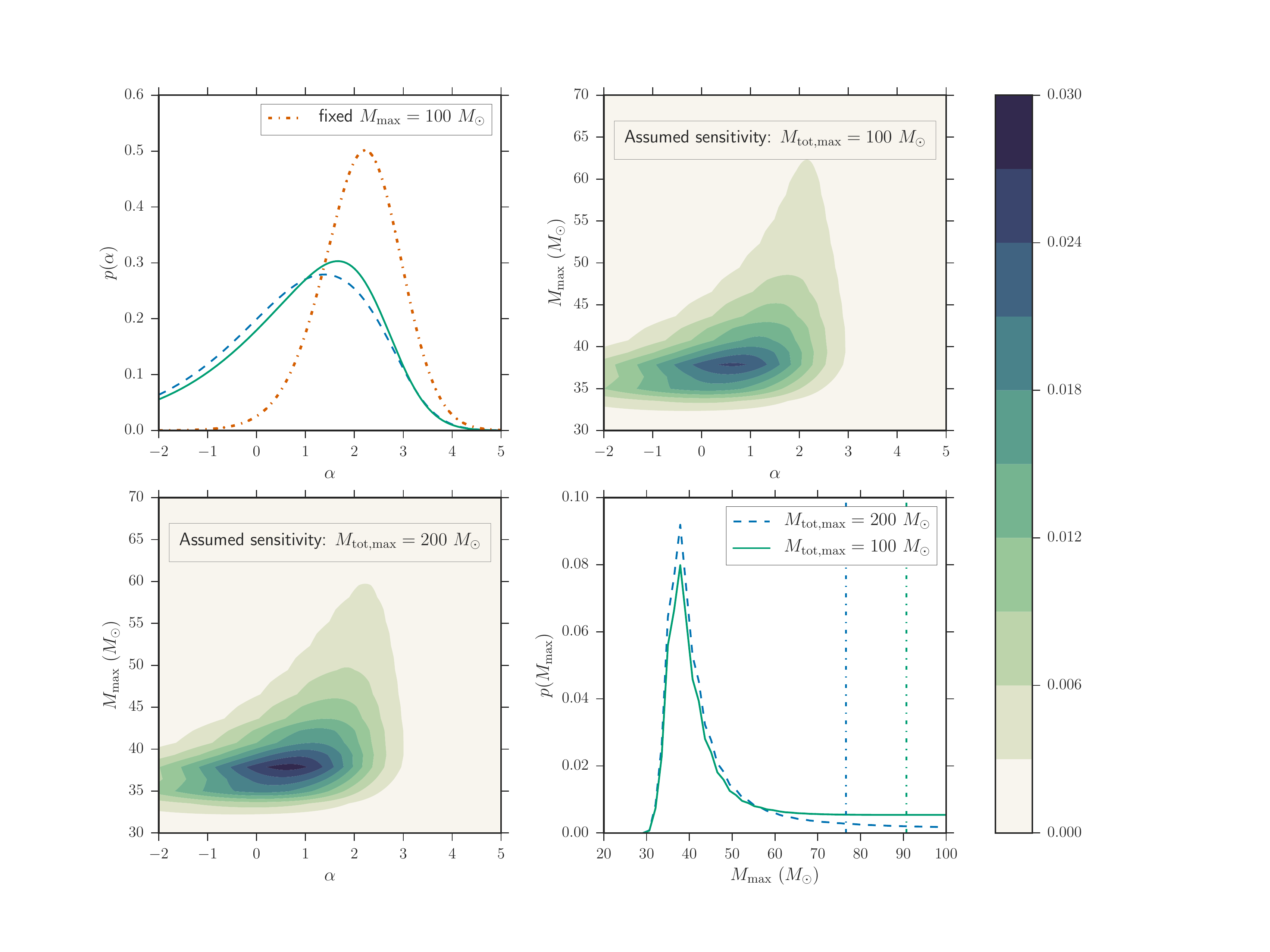}
\caption{Joint fits for $\alpha$, $M_\mathrm{max}$ from the first four LIGO detections. Lower left panel: The posterior PDF $p(\alpha, M_\mathrm{max})$ under the conservative assumption that $M_\mathrm{tot,max} = 100\,M_\odot$. Upper right panel: The posterior PDF $p(\alpha, M_\mathrm{max})$ assuming full matched-filter sensitivity up to $M_\mathrm{tot,max} = 200\,M_\odot$. Upper left panel: The marginal posterior PDF for $\alpha$ under each assumption of $M_\mathrm{tot,max}$ (green solid and dashed blue curves). The orange dash-dotted curve shows the results of the canonical analysis (Eq.~\ref{Eq:LIGOpl}) in which $M_\mathrm{max}$ is fixed at $M_\mathrm{max} = 95\,M_\odot$ and $M_\mathrm{tot, max} = 100\,M_\odot$. Lower right panel: The marginal posterior PDF for $M_\mathrm{max}$ under each assumption of $M_\mathrm{tot,max}$. The vertical dotted lines denote $95 \%$ credible intervals. Throughout, we take a uniform prior on $\alpha$ in the range $-2 \leq \alpha \leq 7$ and on $M_\mathrm{max}$ in the range $M_\mathrm{max} < 100\,M_\odot$, as described in the text.}
\label{fig:palpha-mmax-real4}
\end{figure*}

The results of the joint power law--maximum mass analysis for the set of four detected BBHs is shown in Fig.~\ref{fig:palpha-mmax-real4}. We compute the joint likelihood twice: once fixing $M_\mathrm{tot,max} = 100\,M_\odot$, so that the population of stellar mass BBHs is restricted to total masses $M_\mathrm{tot} \leq 100\,M_\odot$ regardless of $M_\mathrm{max}$ (top right panel) and once fixing $M_\mathrm{tot,max} = 200\,M_\odot$, so that the maximum total mass of the population is allowed to extend to $M_\mathrm{tot,max} = 2 M_\mathrm{max}$. We calculate the marginal posterior PDFs of $\alpha$ and $M_\mathrm{max}$ (top left and bottom right panels) under the assumption of a uniform prior on $\alpha$ in the range $[-2,7]$ and a uniform prior on $M_\mathrm{max}$ in the range $[29\,M_\odot,100\,M_\odot]$ ($29\,M_\odot$ is the minimum posterior sample we generated for the primary component mass of GW150914).

It is clear that properly accounting for our uncertainty on $M_\mathrm{max}$ when fitting the power-law mass distribution increases the support for shallow power law slopes which would otherwise be ruled out under the assumption that the mass distribution extends continuously to $\sim100\,M_\odot$. Allowing for freedom in $M_\mathrm{max}$ shifts the preferred values of $\alpha$ to shallower slopes, even allowing for negative $\alpha$, as compared to the canonical analysis that fixes $M_\mathrm{max} = 95\,M_\odot$ (orange dot-dashed curve in Fig.~\ref{fig:palpha-mmax-real4}). Furthermore, the first four BBH detections already start to constrain $M_\mathrm{max}$. The marginal posterior PDF $p(M_\mathrm{max})$ peaks strongly at $M_\mathrm{max} \sim 40$, and the $95\%$ upper limits on the inferred $p(M_\mathrm{max})$ are $76.6\,M_\odot$ if assuming  $M_\mathrm{tot,max} = 200\,M_\odot$ (or $90.7\,M_\odot$ if we conservatively assume $M_\mathrm{tot,max} = 100\,M_\odot$). Taking $M_\mathrm{tot,max} = 200\,M_\odot$ rather than $100\,M_\odot$ allows the detectable BBH population to extend to $2M_\mathrm{max}$, thereby increasing the expected sensitivity to BBHs with primary component masses $M_\mathrm{max}>50\,M_\odot$. Thus, the non-detection of heavy BBHs yields tighter constraints on the inferred $M_\mathrm{max}$ when we assume $M_\mathrm{tot,max} = 200\,M_\odot$, but the peak of the $M_\mathrm{max}$ distribution remains unchanged.

To explore the impact of future detections on the inferred mass distribution, we repeat this analysis for three simulated BBH populations, two with a power law slope $\alpha = 2.35$ and one with a power law slope $\alpha = 1$. One of the $\alpha = 2.35$ populations, as well as the $\alpha = 1$ population, has a mass gap starting at $M_\mathrm{max} = 50\,M_\odot$, while the other $\alpha = 2.35$ population has a mass gap starting at $M_\mathrm{max} = 40\,M_\odot$.

\begin{figure*}
\includegraphics[width=\textwidth]{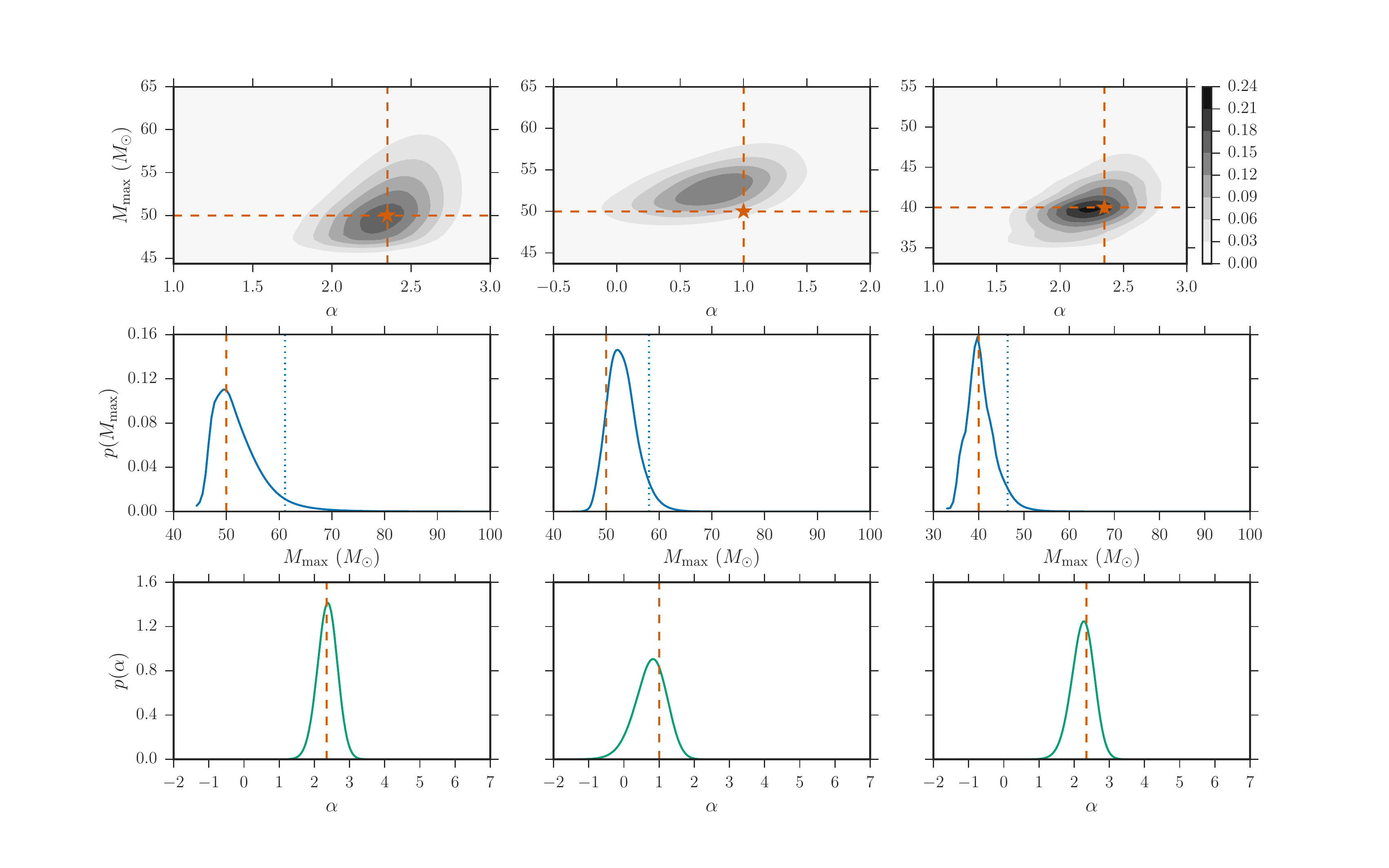}
\caption{Joint fits for ($\alpha$, $M_\mathrm{max}$) using 40 simulated BBH detections from 3 populations, assuming that $M_\mathrm{tot,max} = 2 M_\mathrm{max}$ and that LIGO/Virgo is sensitive up to $M_\mathrm{tot,max} = 200\,M_\odot$. Each column represents a different simulated population where the true $\alpha$, $M_\mathrm{max}$ values are shown by the orange star. Left column: $\alpha = 2.35$, $M_\mathrm{max} = 50\,M_\odot$; middle column: $\alpha = 1$, $M_\mathrm{max} = 50\,M_\odot$; right column: $\alpha = 2.35$, $M_\mathrm{max} = 40\,M_\odot$. The top row shows the posterior PDF $p(\alpha, M_\mathrm{max})$ as recovered from 40 events, the second row shows the marginal PDF of $M_\mathrm{max}$, and the bottom row shows the marginal PDF of $\alpha$ for each simulated population.}
\label{fig:palpha-mmax-40e-wmarginals}
\end{figure*}

The results of this calculation, assuming sensitivity up to $M_\mathrm{tot,max} = 200\,M_\odot$, are shown in Fig.~\ref{fig:palpha-mmax-40e-wmarginals}, where each column corresponds to 40 detections from one of the three simulated populations. We see that 40 detections yield strong constraints on both the slope and maximum mass of the population. If the true population has a cutoff at $M_\mathrm{max} = 40\,M_\odot$ (right column in Fig.~\ref{fig:palpha-mmax-40e-wmarginals}) rather than $50\,M_\odot$, we get tighter constraints on $M_\mathrm{max}$. Similarly, we expect better constraints on the maximum mass for shallower mass distributions, as the non-detection of heavy BHs is more striking for shallow mass distributions (see Fig.~\ref{fig:VTratio-pl}). The maximum mass is indeed better constrained for the population with power law slope $\alpha = 1$ (middle column) than for the population with the same mass cutoff $M_\mathrm{max} = 50\,M_\odot$ but steeper power law slope $\alpha = 2.35$ (left column). As the first four LIGO detections currently favor $M_\mathrm{max} < 50\,M_\odot$ and moderately shallow power law slopes $\alpha < 3$ (Fig.~\ref{fig:palpha-mmax-real4}), we expect that fewer than 40 detections will strongly constrain $M_\mathrm{max}$.

\section{Discussion}
\subsection{Effect of Redshift Evolution}
We have assumed that the merger rate, as measured in the source-frame, is uniform in comoving volume (Eq.~\ref{eq:VTm1m2}). In reality, the merger rate per comoving volume is expected to increase with redshift until $z \sim 2$ \citep[see, for example, Extended Data Fig. 4 in][]{Belczynski:2016a}. This would mean that we have underestimated the $VT$ factors for high mass systems, because high mass systems are detectable at higher redshifts. Thus, we have also underestimated the terms $\langle VT \rangle_\mathrm{low}/\langle VT \rangle_\mathrm{high}$ displayed as a function of power law slope $\alpha$ in Fig.~\ref{fig:VTratio-pl}, and the tension between certain power law slopes and the non-detection of heavy BHs is in fact greater than we predicted. If we assumed a steeper redshift evolution of the merger rate, fewer detections would resolve the mass gap at high confidence.

\subsection{Distribution of mass ratios}
In fitting the mass distribution of BBHs (Eq.~\ref{eq:pl_cutoff}), we assumed that the distribution of mass ratios, $q$, is uniform in the allowed range ${M_\mathrm{min}}/{m_1} < q < \mathrm{min}(M_\mathrm{tot,max}/m_1-1,1)$. In particular, we assumed that for a given primary component mass, $m_1$, the marginal distribution of $m_2 = q m_1$ is given by Eq.~\ref{Eq:LIGOpm2}. However, many BBH formation models predict a preference for equal-mass mergers \citep{DominikI,Rodriguez:2016}. To explore the effects of our assumed mass ratio distribution, we generalize Eq.~\ref{Eq:LIGOpm2}:
\begin{equation}
\label{Eq:pm2-mod}
p(m_2 \mid m_1) \propto \frac{m_2^k}{\mathrm{min}(m_1,M_\mathrm{tot,max}-m_1)^{k+1}-M_\mathrm{min}^{k+1}},
\end{equation}
so that $k=0$ reduces to Eq.~\ref{Eq:LIGOpm2} while $k>0$ favors more equal mass ratios. We find that the choice of $k \geq 0$ does not noticeably impact our results, and we recover consistent posteriors on $(\alpha, M_\mathrm{max})$ if we fix $k=6$ rather than $k=0$. However, we note that there is currently no evidence that the distribution of mass ratios, $p(q \mid m_1)$, deviates from the assumed uniform distribution. Although all of the events so far are consistent with mass ratios close to unity, this is not surprising given the selection effects that favor more equal-mass systems. For a fixed primary mass, $m_1$, assuming full matched-filter sensitivity, we would expect five detections with $q > 0.5$ for every detection with $q < 0.5$, and two detections with $q > 0.7$ for every detection with $q < 0.7$, even if we take $q$ to be uniform in the range $[0,1]$ rather than $[{M_\mathrm{min}}/{m_1},1]$. We can explicitly check if the data favors $k>0$ if we incorporate Eq.~\ref{Eq:pm2-mod} into the power law mass distribution, so that Eq.~\ref{eq:pl_cutoff} becomes:
\begin{equation}
\label{Eq:pl_cutoff_k}
p\left( m_1, m_2 \mid \alpha, M_\mathrm{max} \right) \propto  \frac{m_1^{-\alpha}m_2^k \ \mathcal{H}(M_\mathrm{max} - m_1)}{\min(m_1,M_\mathrm{tot,max}-m_1)^{k+1} - M_\mathrm{min}^{k+1}}.
\end{equation}
We fit the above Eq.~\ref{Eq:pl_cutoff_k} for $k$, marginalizing over $\alpha$ and $M_\mathrm{max}$, and find that, for the first four LIGO detections, the likelihood $p(\mathbf{d} \mid k)$ peaks mildly at $k=0$, but is very broad. Thus, the first four BBHs mildly favor a uniform distribution of mass ratios. Future detections will continue to test this assumption. 

\subsection{Extending to non-power-law mass distributions}
Although a power law provides a good fit to the mass distribution of massive stars, there are theoretical indications that the masses of BHs in merging binaries may diverge from a power-law distribution. For example, supernova theory suggests that there is a nonlinear relationship between the initial zero-age main sequence mass of star and its resulting BH \citep{Belczynski:2016a,Spera:2016}. In fact, PPISN and PISN are associated with significant mass loss and may cause a deviation in the BH mass distribution at masses $>30\,M_\odot$. Additionally, several models predict that a mass-dependent merger efficiency causes the mass distribution for merging binaries to differ significantly from the BH mass function \citep{OLeary:2016}. While we have focused solely on power law fits to the mass distribution, an increased sample of BBH detections will allow us to explore more complicated parametric and non-parametric models and select a model for the mass distribution that best fits the data. Regardless of the model, it is straightforward to include a free parameter (in our case, $M_\mathrm{max}$) that fits for the bottom edge of the upper mass gap.

\subsection{Are there BBHs beyond the gap?}
So far we have restricted our attention to the bottom edge of the upper mass gap, but LIGO is also probing the upper edge of the mass gap in the IMBH search, with results from the first observing run presented in \citet{Abbott:IMBH}. It is theoretically unclear whether BHs exist on the other side of the mass gap (predicted at $\sim 135\,M_\odot$), as the frequency of sufficiently high mass stars is unknown \citep{Belczynski:2016a}. Before accounting for PPISN or PISN, previous population-synthesis predictions placed the maximum BH mass at $80$--$135\,M_\odot$ for zero age main sequence masses $M_\mathrm{ZAMS} <  150\,M_\odot$ \citep{Belczynski:2016,Eldridge:2016,Spera:2016}. However, stars with $M_\mathrm{ZAMS} \gtrsim 200\,M_\odot$ in a sufficiently low-metallicity environment $(Z \sim 0.07 Z_\odot)$ are expected to directly collapse to BHs with masses $\sim 120$--$280\,M_\odot$ \citep{Spera:2017}. We find that LIGO is approximately five times more sensitive to a population of equal mass BBHs just above the mass gap (with total masses in the range $270$--$300\,M_\odot$) than to equal mass BBHs with total masses in the range $10$--$65\,M_\odot$. In the unlikely scenario that a power law continues unbroken over the entire mass range $10 \leq M_\mathrm{tot} \leq 300\,M_\odot$, we could constrain the existence of BBHs above the mass gap by extrapolating the power-law fit from the mass distribution below the gap. If we take a power law with slope $\alpha = 1$, the expected number of detected BBHs below the gap ($5 \lesssim m_1 \lesssim 40$) is $\sim 1.76$ times greater than the expected number of detected binaries directly above the gap ($135 \leq m_1 \leq 150\,M_\odot$). This means that within $\sim 10$ BBH detections with $5 \lesssim m_1 \lesssim 40$, the non-detection of BBHs above the gap would imply that the power-law extrapolation with $\alpha = 1$ breaks down or that BBHs above the gap do not exist. However, if we extrapolate a power law with slope $\alpha = 2.35$ across this mass range, the expected number of detections below the gap is $\sim22.9$ times the expected number of detections directly above the gap, so it would take $\gtrsim 60$ BBH detections to invalidate the power law extrapolation to the other side of the gap. The existing sample of BBHs is too small to place interesting constraints on the existence of systems beyond the gap.

\section{Conclusion}
We have shown that given LIGO's extremely high sensitivity to BBHs with component masses $40 \leq m_1 \leq 100\,M_\odot$, it is statistically significant that the first four detections have been less massive than $40\,M_\odot$. We present a two-parameter model for the BBH mass distribution, consisting of a power law with slope $\alpha$ and a cutoff at $M_\mathrm{max}$, and find that the first four detections already provide evidence for a cutoff to the mass distribution at $M_\mathrm{max} \sim40\,M_\odot$. This cutoff may be the lower edge of a PPISN/ PISN upper mass gap. Furthermore, LIGO-Virgo have recently announced two more BBH detections (GW170608 and GW170814), both of which are less massive than $40\,M_\odot$ and only strengthen our conclusions \citep{Abbott:GW170814,Abbott:GW170608}. We find that within $\mathcal{O}(10)$ BBH detections, the location of the bottom edge of the upper mass gap will be significantly constrained. Our model assumes that all BBHs belong to a single population described by the same power law, so that the detection of a binary with mass in the mass ``gap'' would reset the lower edge of the gap beyond the mass of the newly-detected binary. However, we expect to quickly converge on the true maximum mass of the population within $\lesssim 40$ detections. At this point, the detection of a binary in the mass gap will be statistically inconsistent with this single population, and may indicate a subpopulation of BBHs that did not form directly from stellar-collapse (e.g. primordial BHs or BHs formed through previous mergers). The BBH spin distribution will provide further constraints on the existence of these subpopulations and will allow us to measure the fraction of BBHs that formed through previous mergers \citep{Fishbach:2017,Gerosa:2017}.

\acknowledgments
We thank Will Farr for valuable discussions. MF was supported by the NSF Graduate Research Fellowship Program under Grant No. 1746045. MF and DEH were partially supported by NSF CAREER grant PHY-1151836 and NSF grant PHY-1708081. They were also supported by the Kavli Institute for Cosmological Physics at the University of Chicago through NSF grant PHY-1125897 and an endowment from the Kavli Foundation. DEH thanks the Niels Bohr Institute for its hospitality while part of this work was completed, and acknowledges the Kavli Foundation and the DNRF for supporting the 2017 Kavli Summer Program.

\bibliographystyle{yahapj}
\bibliography{references}

\end{document}